\documentclass[american,aps,prx,superscriptaddress,reprint]{revtex4-1}

\usepackage{xcolor}
\usepackage{graphicx,color}  
\usepackage{wrapfig}
\usepackage{dcolumn}   
\usepackage{bm}        
\usepackage{amssymb}   
\usepackage{amsmath} 
\usepackage{subfig}
\usepackage{float}
\usepackage[unicode=true,
 bookmarks=true,bookmarksnumbered=false,bookmarksopen=false,
 breaklinks=false,pdfborder={0 0 0},pdfborderstyle={},backref=false,colorlinks=true]
 {hyperref}
\hypersetup{
 pdfborderstyle={},allcolors=blue}

\begin{document}

\title{Implementation of a Quantum Engine Fuelled by Information}

\author{John P. S. Peterson}

\thanks{e-mail:  johnpetersonps@hotmail.com}

\affiliation{Institute for Quantum Computing and Department of Physics and Astronomy,
University of Waterloo, Waterloo N2L 3G1, Ontario, Canada}
                        
\author{Roberto S. Sarthour}

\affiliation{Centro Brasileiro de Pesquisas Físicas, Rua Doutor Xavier Sigaud 150, Rio de Janeiro – RJ, 22290-180, Brazil}

\author{Raymond Laflamme}

\affiliation{Institute for Quantum Computing and Department of Physics and Astronomy,
University of Waterloo, Waterloo N2L 3G1, Ontario, Canada}

\affiliation{Perimeter Institute for Theoretical Physics, 31 Caroline Street North, Waterloo, Ontario, N2L 2Y5, Canada}

\affiliation{Canadian Institute for Advanced Research, Toronto, Ontario M5G 1Z8, Canada}
\date{\today}

\begin{abstract}
Here, we show the implementation of a complete cycle of a quantum engine fuelled by information. This engine is a quantum version of the Szilard engine, where information is used to extract heat from the environment and fully convert it into work. In our experiment, this work is used to make a weight, initially in the ground state, reach its excited state. We measure the energy and the state of each component of the engine, after each step of the cycle, and compare them with the theoretical prediction to show that the cycle is implemented with high precision. We also perform experiments to show that the engine is well isolated from the environment after the heat extraction, and we measure the entropy of the weight to show the full conversion of heat into work. Thus, we successfully demonstrate that information can be used as a fuel for single-reservoir engines.
\end{abstract}

\maketitle
The link between thermodynamics and information theory was made more than one hundred years ago, when Maxwell was studying an apparent violation of a thermodynamics law \cite{livroma}. In 1929, the bond between the two areas got stronger when Szilard described an engine fuelled by information \cite{szi}. Some years later, this link was further  explored in studies about  the thermodynamic cost of erasing information \cite{land0,land1,land2,land3,land4}, the definition of the information entropy \cite{entro1,entro2}, the development of the thermodynamics of information and quantum thermodynamics \cite{inft1,inft2,inft3,inft4,inft5}. Although there are some studies about the engine fuelled by information \cite{inft2,sz1,sz2,sz3,sz4,sz5,sz6}, until now it has not been successfully implemented at the quantum level to perform a closed cycle. Here, we show the implementation of a complete cycle of a quantum version of this engine, where information is used to extract energy from the environment. This energy is used to make a system, initially in the ground state, reach its excited state.

According to the second law of the thermodynamics, enunciated by Kelvin, if a thermal engine operates in a closed cycle, the heat extracted from a reservoir cannot be fully converted into work \cite{reif}. This way, most of the thermodynamics engines used in our society (like cars and air planes) are composed of two thermal reservoirs at different temperatures, one cold and one hot. During the cycle of an engine, part of the heat extracted from the hot reservoir is converted into work, and the rest goes to the cold reservoir. Generally, in these engines, the cold reservoir is the environment and the hot reservoir is created, for example, with a combustion process. Commonly, the petrol or other non-renewable energy source is the fuel to create the hot reservoir. It is a well-known fact that a fuel based in petrol can be harmful to the environment.

\begin{figure}[t!]%
    \includegraphics[width=8.5cm]{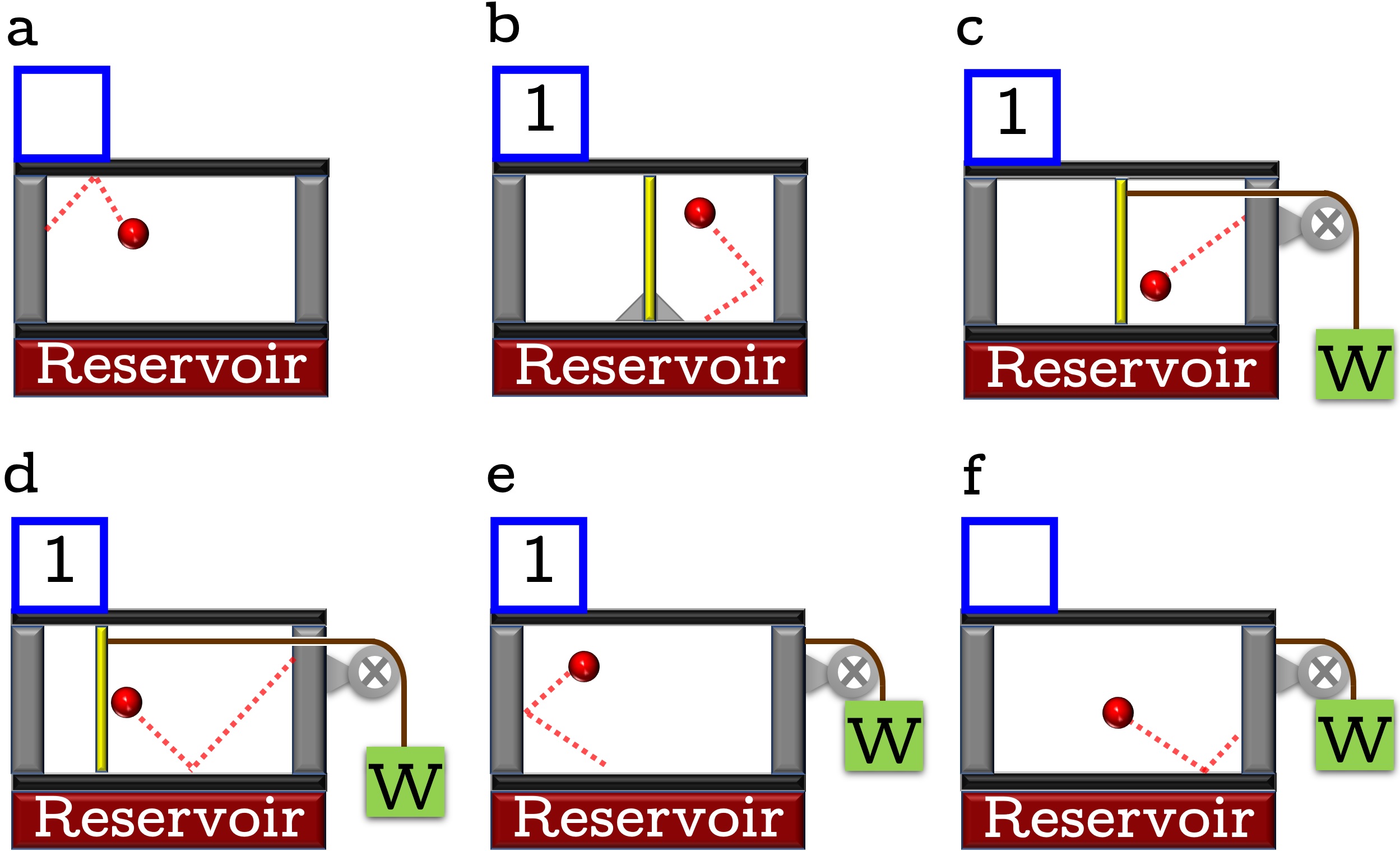}
    \centering
    \caption{Classical Szilard engine - The engine is composed of a container with only one particle (red sphere), a thermal reservoir at temperature $T$, the demon's memory (blue box), and a weight (green box). (\textbf{a}) In the beginning of the cycle, the particle thermalizes with the reservoir. (\textbf{b}) A movable partition is placed inside the container and divides it into two equal parts. The demon measures the particle's position, and stores 0 or 1 in its memory if it is on the left or right side, respectively. (\textbf{c}) Based on the information about the particle's position, the demon chooses the correct side of the container to connect a device to lift the weight. (\textbf{d}) The particle will collide with the movable partition, and all the heat extracted from the reservoir during this process will be fully converted into work to lift the weight. (\textbf{e}) At the end of this process, the weight was lifted but the memory still has the information about the position of the particle. (\textbf{f}) The cycle is closed when this information is erased from the demon's memory.}%
    \label{fig:szcla}%
\end{figure}

A fact that is not well known, and that appears as a violation of the second law, is that the information about the engine's particles can be used to make a closed cycle where the heat extracted from a reservoir is fully converted into work. Surprisingly, this type of engine was proposed in 1929 by Szilard \cite{szi}, and until now it has not been implemented at the quantum level to perform a complete cycle. The Szilard engine is a reformulation of the enigma presented by Maxwell, where an intelligent being (the demon) could use information about the state of a system to apparently violate the second law of the thermodynamics \cite{livroma}. This engine is composed of a container containing only one particle; a demon that measures the particle position and stores the information in its memory; a thermal reservoir at temperature $T$ from which heat is extracted; and a weight that is lifted using the work produced during the cycle. The cycle of this engine is described in Fig. \ref{fig:szcla}. To explain the non-violation of the second law, Szilard pointed that the demon's memory needs to be erased to close the engine cycle, and he postulated that some energy needs to be spent when the demon performs the measurement. Some years later, Zurek, Bennett and Penrose \cite{sz6,lad1,lad2} used the Landauer's principle to explain the non-violation. According to this principle, at least $k_{B}Tlog(2)$ of energy needs to be spent to erase one bit of information from a memory \cite{land0}. Here, $T$ is the temperature of the reservoir used in the erasing process and $k_{B}$ is the Boltzmann constant. Since the container with the particle is divided into two parts (see Fig. \ref{fig:szcla}), the demon only needs to store one bit of information in its memory after measuring the position of the particle. Because of Szilard's work, today we know that instead of petrol or other non-renewable energy source, the information can be used as a fuel for engines with only one reservoir. Here, our goal is to implement a complete cycle of this engine at the quantum level, showing that information can be used as a fuel.

\begin{figure}[t!]%
    \includegraphics[width=8.5cm]{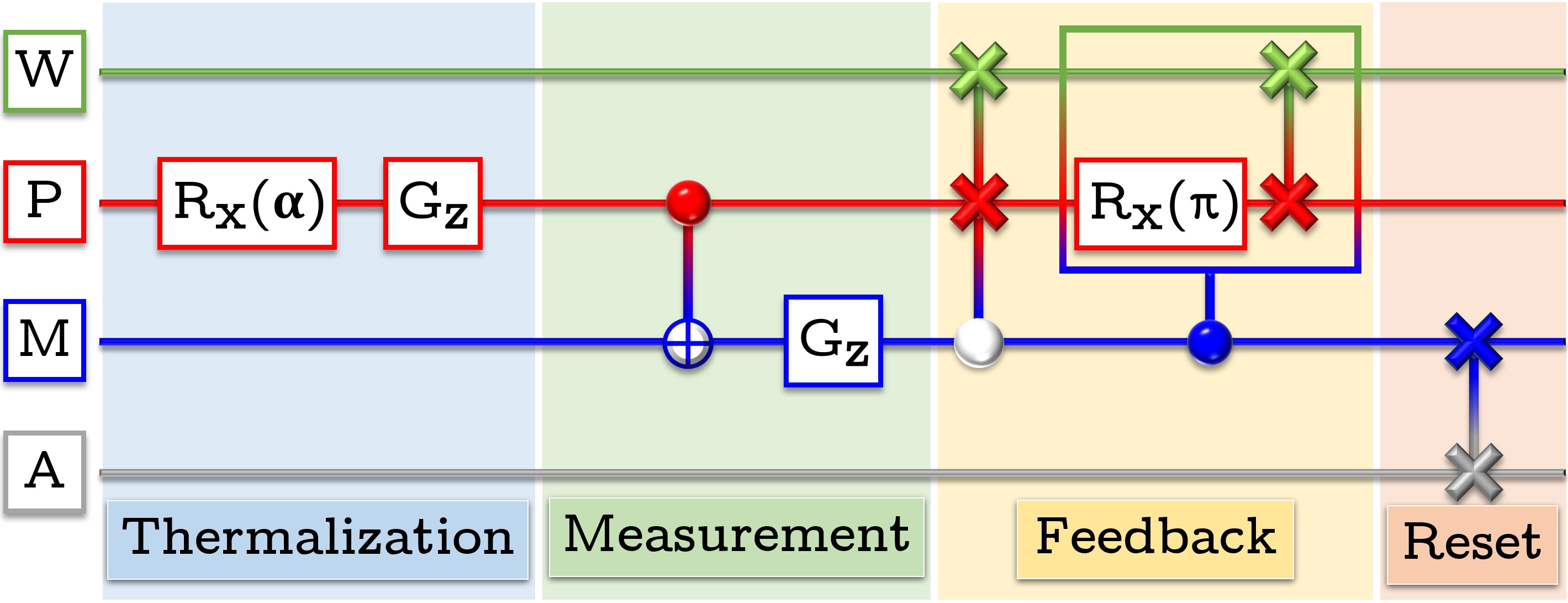}
    \centering
    \caption{Quantum Szilard engine - The qubits W, P, M and A represent the weight, the particle, the demon's memory and the ancillary, respectively. $\textrm{R}_{\textrm{x}}(\theta)$ is a rotation by an angle $\theta$ around the x-axis. Knowing that the particle starts in its ground state, we can use Eq.~\eqref{eq:tstate} to write $\alpha$ as a function of the reservoir temperature, $\alpha(T_{x})= 2 \arccos \left [  \left ( 1+e^{-\hbar \omega/k_{B}T_{x}} \right )^{-1/2}\right ]$.  $\textrm{G}_{\textrm{z}}$ is a magnetic field gradient along the z direction that is applied to a system to remove the non-zero order coherence terms of its state \cite{grad}. In the circuit, we have three controlled gates: a CNOT \cite{livronc} (in the measurement), a controlled Swap \cite{livronc} (first gate of the feedback) and a composition of controlled rotation and Swap (second gate of the feedback). For these gates, if the sphere in the control qubit is white, an operation will be applied to the target qubits when the control qubit is in the excited state, but if it has another color, the operation will be applied when the control qubit is in the ground state. Then, only the states of the target qubits will be changed by a controlled gate. The ancillary and the memory start the cycle with the same initial state. Thus, to reset the memory, a Swap gate is applied to exchange the states of the memory and the ancillary.}%
    \label{fig:szqua}%
\end{figure}

In the quantum scenario, the Szilard engine can be implemented using four systems of two energy level \cite{sz4}, that represent physically a four-qubits system. One qubit represents the weight, one is the particle, the other is the demon and its memory, and the last is an ancillary qubit used to erase the demon's memory. The cycle of the engine is described in the quantum circuit shown in Fig. \ref{fig:szqua}. It is divided into four steps: an initial thermalization where, on average, the particle extracts heat from a reservoir; a measurement for the demon to obtain information about the particle state; a feedback process where information is used to fully convert heat into work and put the weight in its excited state; and the erasing process to close the cycle.

\begin{figure*}[t!]%
 \includegraphics[width=18.0cm]{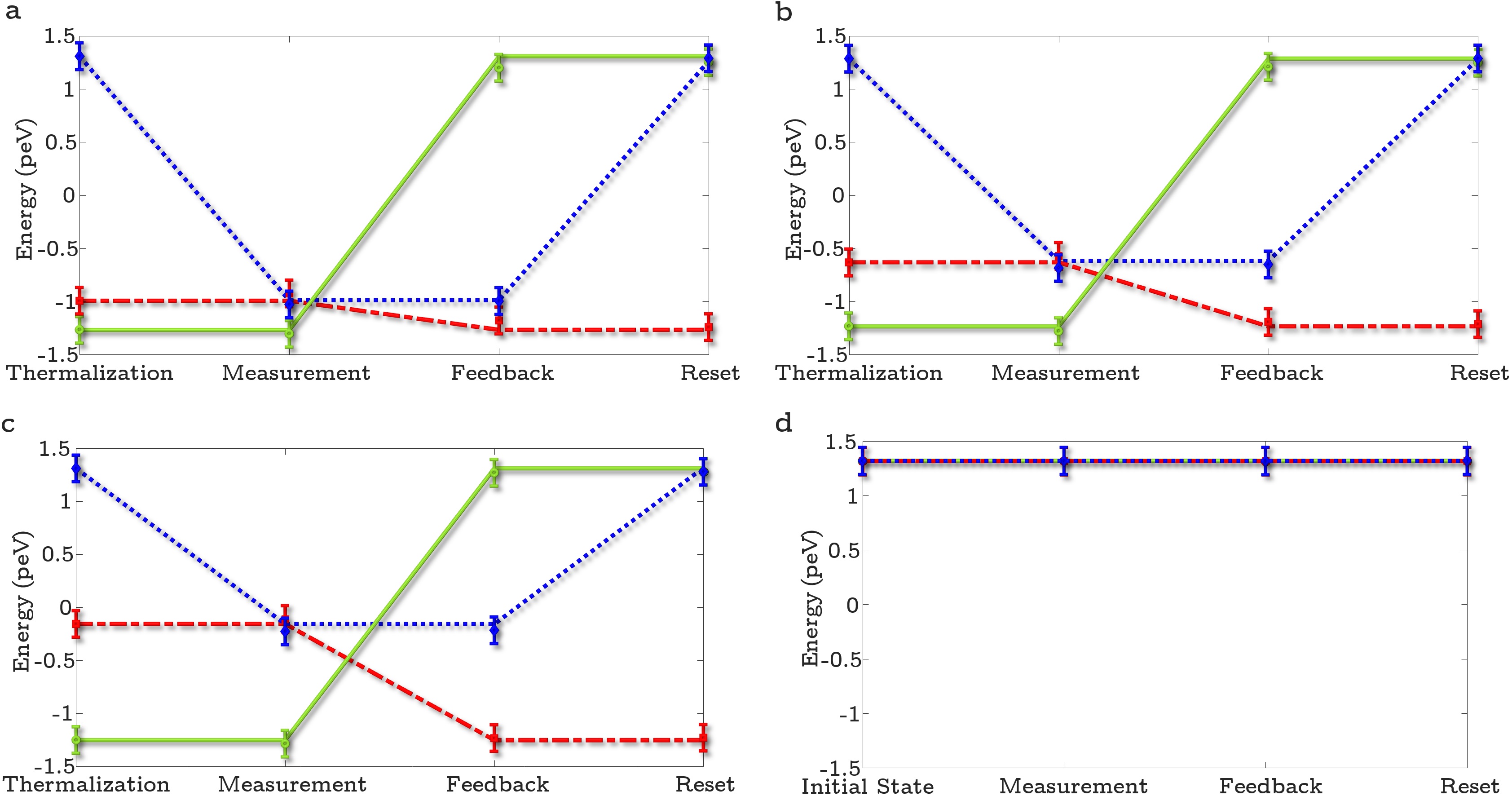}
    \centering
    \caption{Average energy - The lines represent the theoretical prediction and the points are the experimental values of the average energy of each system after every step of the cycle. The particle, weight and memory are represented by the colors red, green and blue, respectively. The Hamiltonian of the particle ($\mathcal{H}_{P}$), weight ($\mathcal{H}_{W}$), and memory ($\mathcal{H}_{M}$) are equal to $\omega \hbar \sigma_{z}$, with $\omega = 2000$ rad/s. The particle was in the ground state before the thermalization with a reservoir at temperature (\textbf{a}) $k_{B}T_{1} = 1.33$ peV, (\textbf{b}) $k_{B}T_{2} = 2.51$ peV, and (\textbf{c}) $k_{B}T_{3} = 10.91 $ peV. In (\textbf{d}), all the systems start the cycle in the excited state and the thermalization is not performed. In our experiments, errors in the implementation of pulses and gradients \cite{pulso,grad}, some random errors in the energy measurement, and the relaxation process \cite{livro} are the main sources of errors. The error bars of 0.1 peV were determined considering these errors in a Monte Carlo algorithm for estimating errors, which is described and employed in Ref. \cite{ter2,ter3}.}%
    \label{fig:result}%
\end{figure*}

 The nucleus of the $^{13}\textrm{C}$ atom has a spin 1/2 and, under a static magnetic field, it will be a system with two energy levels that physically represents a qubit \cite{livro}. To implement the Szilard engine, following the quantum circuit shown in Fig. \ref{fig:szqua}, we used the four nuclear spins of the carbon atoms of the $^{13}\textrm{C}$-labelled transcrotonic acid molecule to physically represent our four qubits. The processes described in the quantum circuit, the quantum gates shown in Fig. \ref{fig:szqua}, were implemented using the nuclear magnetic resonance (NMR) technique. In this technique, the state of the qubits can be controlled by radio-frequency pulses and magnetic field gradients, and can be determined measuring the magnetization of the nuclear spins. More details about this technique and the system employed are presented in the Methods section. Recently, this technique was used to study some key components present in the Szilard engine \cite{land4,ter1,ter2,ter3}.

The experiment was performed using a Bruker Avance III $700$ MHz NMR spectrometer. To implement the processes with high precision, we used the algorithm presented by Peterson et al. \cite{pulso} to optimize the amplitude and phase of the radio-frequency pulses. The initial state of the system was prepared using the algorithm presented by Peterson et al. \cite{grad}. To quantify the precision of our experiments, we determined the state of each qubit, after every step of the cycle, using quantum state tomography \cite{livronc}. Then, we compared the experimental states with the theoretical ones using the quantum fidelity \cite{livronc}, $\mathcal{F} = tr[\sqrt{\sqrt{\rho_{exp}} \rho_{teo} \sqrt{\rho_{exp}}}]$, where $ \rho_{exp} $ and  $\rho_{teo}$ are the experimental and theoretical states, respectively. The quantum fidelity can have a value in the range $[0,1]$, and will be 1 only if $ \rho_{exp}  = \rho_{teo} $. In our experiments, the quantum fidelity is greater than 0.999 for all states. This means that the theoretical and experimental states are very close and, consequently, the errors during the implementation of the quantum engine cycle are small.

We implemented the Szilard engine with four different configurations. In the first three, the weight and the particle begin in their ground state, and the memory begin in its excited state. To show that the engine works for any temperature of the reservoir, we change this temperature in each implementation. We used the following temperatures: $k_{B}T_{1} = 1.33$ peV,  $k_{B}T_{2} = 2.51$ peV, and  $k_{B}T_{3} = 10.91 $ peV. After the thermalization with the reservoir, which is performed with the application of a radio-frequency pulse and a gradient of magnetic field in the nuclear spins, the particle will be in the state: 
\begin{equation}\label{eq:tstate}
 \begin{split}
\ \rho_{T_{x}} = \dfrac{e^{-\mathcal{H}_{P}/k_{B}T_{x}}}{Tr[e^{-\mathcal{H}_{P}/k_{B}T_{x}}]},
 \end{split}
 \end{equation}  
where $\mathcal{H}_{P} = \hbar \omega \sigma_{z}$ is the Hamiltonian of the particle, with $\omega = 2000 $ rad/s and $\sigma_{z}$ is the Pauli matrix. In the last implementation, we considered the case without thermalization, and the four qubits starting in their excited states. In theory, in this case, the states of the qubits should not change after the implementation of each step of the cycle. Then, with our experimental results, we can quantify how much energy is leaving the system (composed of the weight, particle and memory) and going into the environment. If this amount of energy is close to zero, we can say that the system is well isolated during the cycle.

For each configuration, we measured the average energy of the weight, the particle, and the memory after each step of the cycle. As shown in Fig. \ref{fig:result}, the energy measured at each step of the cycle agrees with the theoretical prediction. Considering the error bars in Fig. \ref{fig:result}(a-c), we can see that the weight is very close to its excited state after the feedback process, and the memory state has not changed significantly during this process. Thus, the information about the particle state is used to fully convert the heat extracted from the reservoir into work to put the weight in its excited state. Another way to confirm this full conversion is to verify if the entropy variation of the weight is null during the feedback process. For this, we used the weight state to calculate the entropy variation, and we obtained the following results: $\Delta S_{T_{1}} = 0.1$ peV, $\Delta S_{T_{2}} = 0.1$ peV, and $\Delta S_{T_{3}} = 0.2$ peV, with an error of $\pm 0.2$ peV. This show that, considering the error in the entropy and the amount of energy obtained by the weight ($ \approx 2.5$ peV), the amount of entropy produced is very small. At the end, to close the cycle, the demon's memory is erased and the average energy spent during this process is equal to: $E_{T_{1}} = 2.3 \pm 0.1 $ peV,  $E_{T_{2}} = 1.9 \pm 0.1 $ peV and $E_{T_{3}} = 1.5 \pm 0.1 $ peV, see Fig. \ref{fig:result}(a-c). The amount of energy spent to erase the memory is equal to the energy that decreased from the demon's memory during the measurement process.

 To confirm the accuracy of our results, besides comparing the measured energies with the theoretical prediction, we performed quantum state tomography to determine the state of each qubit in every step of the cycle, and compared it with the theoretical one using the fidelity, which was always greater than 0.999. We also verified the entropy variation of the weight state to show that most of the energy obtained by the weight is work. Finally, we implemented the process with the whole system starting in its excited state, and we could see that the energy that leaves the system is very small, see Fig. \ref{fig:result}(d). This way, we can affirm that we finally managed to implement with high precision an engine that was presented almost one hundred years ago by Leo Szilard. In our quantum version of the Szilard engine, the nuclear spins of the $^{13}\textrm{C}$ atoms of the $^{13}\textrm{C}$-labelled transcrotonic molecule are controlled with the NMR technique to perform the engine cycle. In this engine, part of the energy that makes the weight reach the excited state is the heat from the thermal reservoir (extracted from the particle during the feedback), and the other part is the energy that leaves the memory when the measurement was performed. Thus, the information obtained during the measurement is the fuel for this engine to extract energy from the environment. To erase the memory and close the cycle, on average, we have to spend an amount of energy equal to the one obtained during the measurement process. We believe that our experimental study will contribute to make the engines of our society start to use information as a fuel. This way, the future engines with a single reservoir will extract energy from the environment and no fuel based in petrol or non-renewable energy source will be required.

\section{Methods}

\begin{figure}[h!]%
    \includegraphics[width=8.5cm]{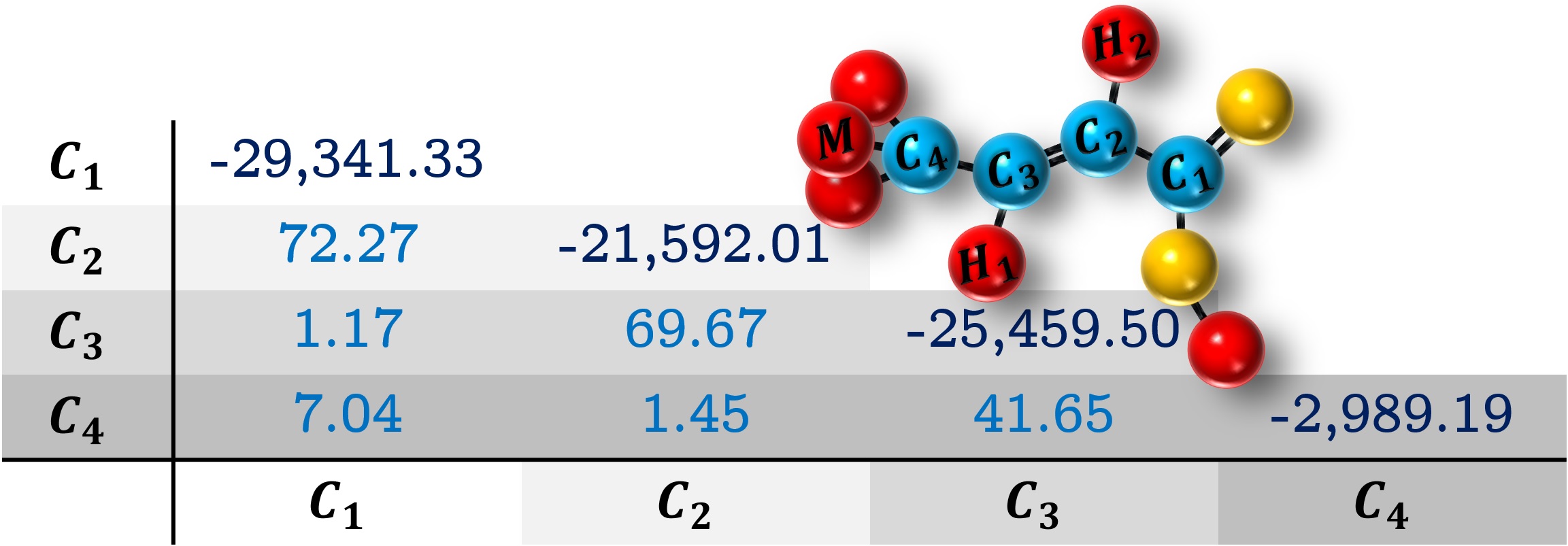}
    \centering
    \caption{$^{13}\textrm{C}$-labelled transcrotonic acid molecule - The carbon, hydrogen and oxygen atoms of the transcrotonic acid molecule are represented by the blue, red and yellow spheres, respectively. The non-diagonal terms in the table are the $J$ coupling constants of the $^{13}\textrm{C}$ nuclear spins of the molecule. The chemical shifts \cite{livro} of the nuclear spins are the diagonal terms. The values in the table are in Hertz.}%
    \label{fig:molecula4q}%
\end{figure}

We used a liquid state sample of $^{13}\textrm{C}$-labelled transcrotonic acid dissolved in acetone. The four $^{13}\textrm{C}$ nuclear spins in this sample physically represent a four-qubit system. In the NMR technique, the sample is placed in a strong and uniform magnetic field in the z-direction. For our sample, the natural dynamics can be described, with a good approximation, by the Hamiltonian:
\begin{equation}\label{eq:h0}
 \begin{split}
\ \mathcal{H}_{0} = \sum_{k}\frac{\hbar(\omega_{k}-\omega_{R})\sigma_{z_{k}}}{2} + \sum_{k \neq n}\frac{\pi\hbar J_{kn}\sigma_{z_{k}}\sigma_{z_{n}}}{4},
 \end{split}
 \end{equation} 
where $\omega_{k}$ and $\sigma_{\beta_{k}}$ are, respectively, the angular oscillation frequency and the Pauli matrix $\beta$ of the $k$-th nuclear spin. $\omega_{R}$ is the angular frequency of the rotating frame and $J_{kn}$ is the scalar coupling constant of the spins $k$ and $n$ \cite{livro}. The values of $\omega_{k}$ and the $J_{kn}$ are shown in Fig. \ref{fig:molecula4q}. In our experiments, we fixed $\omega_{R} = (\omega_{1} + \omega_{4})/2 $.

The nuclear spins states are controlled by radio-frequency pulses applied in the $xy$ plane with an angular frequency $\omega_{R}$. The Hamiltonian that describes the interaction of the spins with a pulse is given by:
\begin{equation}\label{eq:hc}
 \begin{split}
\ \mathcal{H}_{C}(t) = \hbar \Omega(t) \sum_{k}\frac{\cos[ \phi(t) ] \sigma_{x_{k}} + \sin[ \phi(t) ] \sigma_{y_{k}}}{2},
 \end{split}
 \end{equation} 
where $\Omega(t)$ and $\phi(t)$ are the pulse amplitude and phase, which need to be optimized to implement a quantum gate with high precision. We used the method presented by Peterson et. al \cite{pulso} to perform this optimization. Since this new method already takes into account some common sources of errors, and the sequence length ($\approx  0.08$ s) is much shorter than the relaxation times ($T_{1} \approx  10$ s and $T_{2} \approx  1$ s), we do not need to use other techniques to correct errors.

The initial state was prepared following the method presented in Ref. \cite{grad}, which uses magnetic field gradients to produce a pseudo-pure state, starting from a thermal state. The magnetization of each spin was measured, following the quantum state tomography method \cite{livronc}, and used to determine the qubit states. 

\textbf{Acknowledgements}

We thank Ana C. S. Pinto,  Hemant Katiyar,  Frederico Brito, Roberto M. Serra and Marcelo S. Sarandy for valuable discussions that helped to develop the ideas presented in this paper. We acknowledge financial support from CNPq, Ministery of Innovation,
Science and Economic Development (Canada), the Government of Ontario, CIFAR, Mike and Ophelia Lazaridis.

\textbf{Author Contributions}

J.P.S.P. proposed the project and designed the experiments. J.P.S.P. performed the simulations, the experiments and the data analyse. R.L and R.S.S supervised the project. J.P.S.P. and R.S.S wrote the manuscript with feedback from R.L.

\end{document}